# Attacking AI Accelerators by Leveraging Arithmetic Properties of Addition

Masoud Heidary, Biresh Kumar Joardar, Department of ECE, University of Houston

***Abstract*— The dependability of AI models relies largely on the reliability of the underlying computation hardware. Hardware aging attacks can compromise the computing substrate and disrupt AI models over the long run. In this work, we present a new hardware aging attack that exploits commutative properties of addition to disrupt the multiply-and-add operation that forms the backbone of almost all AI models. By permuting the inputs of an adder, the attack preserves functional correctness while inducing unbalanced stress among transistors, accelerating delay degradation in the circuit. Unlike prior approaches that rely on input manipulation, additional trojan circuitry, etc., the proposed method incurs virtually no area or software overhead. Experimental results with two types of multipliers, different bit widths, a mix of AI models and datasets demonstrates that the proposed attack degrades inference accuracy by up to 64% in 4 years, posing a significant threat to AI accelerators. The attack can also be extended to arithmetic units of general-purpose processors.**

## I. INTRODUCTION

As semiconductor technology continues to scale down to nanometer regimes, the reliability of integrated circuits (ICs) is increasingly challenged by various aging phenomena such as Hot Carrier Injection (HCI), Time Dependent Dielectric Breakdown (TDDB), and Negative Bias Temperature Instability (NBTI) [1], [2], [3], [4]. Among these, NBTI has emerged as a critical concern due to its significant impact on PMOS transistors under prolonged negative gate bias [5]. NBTI results in gradual performance degradation, reduced switching speed, and increased power use in ICs. Over time, this degradation can accumulate and cause functional failure, which is particularly concerning in safety-critical systems such as AI, autonomous vehicle and medical devices [6].

While aging is a natural process, hardware attack can accelerate the process, causing unexpected errors resulting in accuracy loss during inferencing. Hardware attacks are malicious activities aimed at exploiting vulnerabilities in the physical ICs to cause damage or extract data. Hardware attacks can include physical tampering [7], side-channel attacks [8], and aging attacks [9]. Here, we focus on aging attacks, where an adversary deliberately accelerates the aging process of specific transistors or circuit components. The aim of the attacker can be to induce premature failure, degrade performance, or create security vulnerabilities over time [10], [11]. In the context of AI accelerators, hardware attacks can result in erroneous computations [12], or they can be used to steal model parameters [13]. Various reliability techniques have been developed to counter aging. However, these methods are often bypassed [14], [15].

Attackers can launch NBTI-based aging attacks using voltage over-scaling, boosting, or thermal variations (e.g. thermal cycling or localized heating) to accelerate NBTI degradation [16]. Workload patterns that stress transistors with frequent switching or high activity can also accelerate degradation [17]. Hardware trojans can manipulate voltage and increase transistor stress, accelerating aging [18]. However, these attacks typically deviate from normal operations and are likely to be detected by anomaly detection methods.

In this work, we develop a new circuit tampering-based hardware aging attack leveraging mathematical properties, to target AI accelerators and related applications. Specifically, the attack manipulates adders within multipliers by leveraging the commutative property of addition. Since multipliers typically consist of arrays of adders, such tampering compromises the multiplier, resulting in incorrect computations. While the attack can be applied to any commutative logic blocks, here we target multipliers due to their widespread use in AI accelerators like systolic arrays, and in ALUs of general purpose processors like CPU, GPU, making it one of the most common and ubiquitous hardware blocks [19], [20]. Moreover, AI models like CNNs, GNNs, transformers, etc., predominantly rely on matrix multiplications, making them particularly vulnerable to proposed attack as we show later. Repeated computations under the proposed attack cause accelerated device wear out, gradually degrading performance and reliability over time, leading to erroneous computations [21].

The key insight behind our attack is that: (a) there is unequal stress among the various PMOS transistors in a multiplier, and (b) addition is commutative (i.e., $A + B = B + A$). Hence, even if we exchange the input connections of a full adder, e.g. $B \leftrightarrow C$, the sum remains constant, i.e., $sum = A \oplus B \oplus C \equiv A \oplus C \oplus B$, and the correctness of the output is not affected. Figure 1 illustrates the idea. As we can see in Figure 1, swapping input pins in a full adder does not alter the output (sum and carry remains unchanged). However, exchanging the input pins changes which transistors experience stress due to negative bias for longer, as we explain later. When done strategically, this exacerbates the unequal stress problem, causing a few chosen transistors to age much faster than expected, thereby slowing the adder and multiplier leading to timing faults, and accuracy loss during AI inferencing.

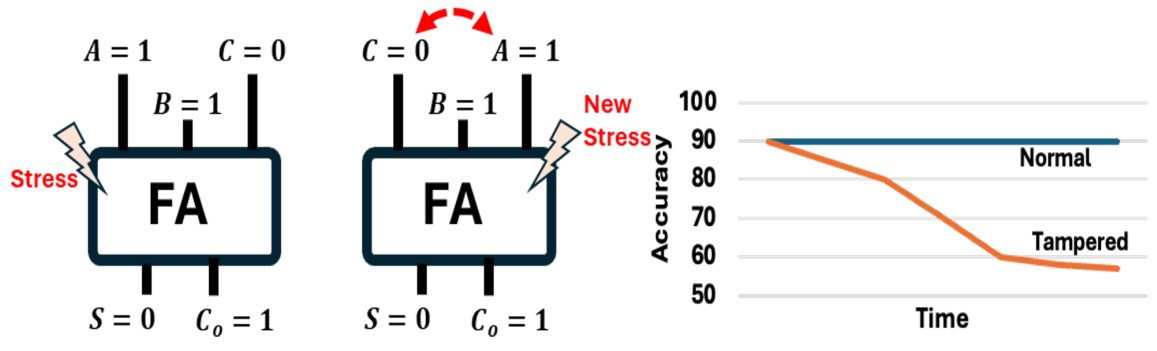

(a) (b) (c)

**Figure 1.** Full Adder with three inputs, (a) normal connection, (b) tampered adder, (c) effect on model accuracy due to tampering.

Following prior work [35][36][37], we assume that the proposed attack is launched either by a malicious foundry or a third-party IP vendor, aiming to induce accuracy loss during inference, disrupt operations and/or damage the manufacturer's reputation. Moreover, the attack has virtually no area overhead, as it only requires rewiring input pins. This leads to minimal structural changes, preserving the original design while strategically increasing stress on transistors in the critical path. Furthermore, the attack is difficult to detect since the output behavior remains logically correct and timing violations only happen over time. Finally, the proposed attack is applicable to a wide variety of architectures that rely on adders/multipliers e.g. CPU, GPU, systolic arrays, etc. Experimental results indicate that the attack causes 51%, 26%, 14% accuracy loss on average for CNNs, GNN, and transformer models respectively after 4-years of operation despite having a static timing guard band.

The rest of the paper is organized as follows. Section II discusses relevant prior works. Section III provides a background discussion on NBTI aging and process variation. Section IV presents the proposed attack, detailing the algorithm for targeting commutative blocks and its application to rewire (tamper) the full adders within the multipliers of a systolic array. Section V describes the experimental setup, and relevant results to assess the effectiveness of the proposed attack. Section VI concludes with a discussion of the key findings.

## II. Prior Work

In this section we present relevant prior work on hardware aging and aging attacks on AI accelerators.

### A. Hardware Aging

Hardware aging is a critical challenge in modern computing systems, affecting both reliability and performance over time [2]. Various aging mechanisms effect transistor characteristics, leading to failure or increased power consumption. Among these, NBTI [1], [5], HCI [3] and TDDB [4] are the most prominent. NBTI mostly affects PMOS transistors, increase their threshold voltage ($V_{th}$) over time, which in turn increases delay and reduces drain current, leading to computational errors [1], [5]. HCI impacts both NMOS and PMOS transistors, causing degradation due to high-energy carriers trapped in the oxide layer, which leads to increased resistance and reduced transistor performance [3]. TDDB on the other hand, occurs when the gate oxide gradually breaks down, leading to leakage currents and eventual failure [4]. The degradation due to aging is further exacerbated by process variation. Due to variations from manufacturing, operating conditions and usage, some transistors age at a faster rate than the others [22], which makes mitigation challenging.

### B. Hardware Aging Attacks

Hardware attack exploits vulnerabilities in physical devices to compromise their security, integrity, or functionality. These attacks can range from probing and tampering to side-channel attacks and fault injection attacks [23]. Aging attacks leverage the natural wear out, to induce failure or degrade performance. NBTI based attacks have been proposed in prior work [5], [24], [25], which accelerate the NBTI process thereby compromising device functionality and reliability over time. These attacks can be implemented during design or manufacturing stages.

NBTI aging attack by tampering workload pattern is proposed in [5]. A self-heating hardware trojan is proposed in [26] that leverages excessive localized heat in advance nodes such as FinFETs and nanowires to gradually shift threshold voltage $V_{th}$ and activate latent malicious behavior. While this attack exploits thermal effects, it is hard to reliably trigger or control in complex logic blocks due to power management policies. Workload aware trojans [25] manipulate NBTI induced degradation using statistical models to align device failure within a desired time window. However, this method relies on static profiling and may not be as efficient under dynamically changing workloads.

System level blocks like network-on-chips have also been targeted using aging attacks [10]. This attack manipulates routing policies to shift traffic patterns and unbalance the aging, to delay the failure till post warranty period. Method in [27] utilizes NBTI aging in ICs as a trigger to activate other malicious circuits. However, it does not implement any new aging attack itself.

### C. Aging Attacks on AI Accelerators

Hardware aging attacks have been shown to cause timing violations and functional errors in AI inferencing [28]. In [29], the authors show that aging induced timing errors in Multiply-Accumulate (MAC) units and weight buffers progressively degrade the classification accuracy. In [30], the authors analyze aging for systolic arrays, and increase their vulnerability towards data corruption. Similarly, in [31], the authors show that aging effects can make AI hardware less safe, especially under long operational lifetimes.

In contrast to these prior works, the proposed attack exploits the inherent properties of mathematical computations to cause errors in MAC units of AI accelerators. Specifically, by exploiting the commutative property of addition, the attack induces accelerated NBTI effect in adders and multipliers of a systolic array. As we show later, the attack results in unequal aging within the systolic array without any noticeable implementation overhead. This leads to degradation in inference accuracy over time considering various AI models.

## III. Background

In this section, we provide a brief background on NBTI aging, process variation along with their mathematical models.

### A. NBTI aging

NBTI is a prominent aging mechanism in PMOS transistors that leads to an increase in threshold voltage ($V_{th}$) over time ($t$). this degradation can be modeled as:

$$\mathrm{V_{th}}(t) = V_{th}(t=0) + |\Delta V_{th}(t)| \qquad (1)$$

The incremental shift $|\Delta V_{th}(t)|$ reduces the drive current, which in turn increases the switching delay of transistors. In high-performance digital circuit, such as AI accelerators, this will cause timing violations and reliability issues, as in MAC units in systolic array [32].

NBTI aging occurs when a PMOS transistor is in the ON state (i.e. $V_{gs} = -V_{dd}$). During this state, holes in the inversion layer interact with pre-existing defects in the silicon-dioxide layer, forming interface traps at the $Si/SiO2$ boundary. These traps capture charge carriers, increasing the threshold voltage and slowing down switching speed. The Reaction-Diffusion model describes this behavior [5], [33]:

$$|\Delta V_{th}(t)| = \left(\frac{\sqrt{K_v^2.\alpha.T_{data}}}{1-\beta(t)^{1/2\lambda}}\right)^{2\lambda} \quad (2)$$

Here, $T_{data}$ represents the duration for which stress is applied, and $\alpha$ is the probability that the PMOS input is at logic '0', resulting in negative bias at PMOS transistor, and $\lambda$ represents the time exponent. Overall, the product $\alpha.T_{data}$ quantifies the effective duration a PMOS is under stress. We refer the reader to [5], [33] for more detail about the NBTI model. Together, Equation (1) and (2) quantify the progressive degradation in $V_{th}$ as a function of input activity and usage time.

### B. Process variation

Process variation refers to statistical fluctuations in device parameters arising from imperfections in semiconductor fabrication processes, such as lithography or doping inconsistencies. These variations lead to deviations in transistor characteristics across the die, including the initial threshold voltage $V_{th}(t = 0)$, which can significantly impact overall circuit timing and reliability [34].

Transistors with initially higher $V_{th}$ may be more susceptible to accelerated degradation due to NBTI. This results in non-uniform aging across gates, increasing the likelihood of timing failure along particular paths. The statistical distribution of threshold voltage deviation due to process variation is modeled using Gaussian distribution denoted as $N$:

$$\Delta V_{th,PV} \sim N\left(0, \sigma_{V_{th}}^2\right) \quad (3)$$

Accordingly, the initial threshold voltage becomes:

$$V_{th}(t = 0) = V_{th,nominal} \pm \Delta V_{th,PV} \quad (4)$$

Combining Equations (1), (2) and (4), the threshold voltage at any future time (t) is expressed as:

$$V_{th}(t) = V_{th,nominal} \pm \Delta V_{th,PV} + |\Delta V_{th}(t)| \quad (5)$$

This formulation allows the incorporation of both spatial variation and temporal degradation of $V_{th}$ in simulation (e.g. Cadence Spectre, OCEAN), enabling accurate estimation of delay, and accuracy for aging aware design analysis.

## IV. Proposed Method

In this section we present the proposed technique in detail. First, we present our underlying assumptions about the attack. Next, we present the attack followed by discussion about how to implement it in complex systems such as a systolic array-based AI accelerator.

**Assumption and Threat Model:** IC manufacturing is a global undertaking, involving multiple vendors from various countries. Here, we assume that the hardware is designed by vendor A, with help from vendor B (an untrustworthy foundry or a third-party IP vendor). Vendor A performs the final post-manufacturing testing before selling to customers. Our threat model assumes that Vendor B (malicious foundry or a third-party IP vendor) is a malicious actor, and they tamper the circuit before fabrication. This threat model aligns with multiple prior work that considers untrusted foundries or third-party IP vendors as adversaries (e.g. [35], [36], [37]). The attacker's objective is to accelerate the degradation of the hardware (MAC unit here) to introduce error during inference without getting detected during the post-manufacturing testing by vendor A.

### A. Attacking One Adder: Overview

While the proposed attack can be launched on any logic block with commutative property, here, we specifically focus on adders and multipliers, due to their ubiquitous nature and their importance for AI workloads. The proposed NBTI-based attack leverages both the inherent imbalance in the switching activity of transistors and the commutative nature of addition.

In a full adder with $A, B$ and $C$ as inputs, the attack is achieved by changing the physical input connections of adders e.g. $B \leftrightarrow C$. Despite its simplicity, the change can create an imbalance in 'stress' among the various transistors in the adder. This imbalance leads to differential aging, where some transistors age faster than others, potentially causing performance degradation and reliability issues in the targeted circuits. Here, following prior work, we define a PMOS as under 'stress' when the voltage difference between the gate and source of the PMOS is negative. This happens when the input gate voltage to the PMOS is $0V$, and the source voltage is greater than $0V$ i.e. when the PMOS is in the 'on' state. This is captured by the term $\alpha$ in Equation (2) where, α is the likelihood that a data is '0' (leading to negative bias).

Figure 2 illustrates the effect of the attack using a single full adder (FA) as a representative commutative logic block. In Figure 2(a), we assume that the input to an adder is '$A = 1, B = 1, C = 0$'. When this input is applied, transistor $T1$ is not under stress as its gate input is logic '1'. Next, Figure 2(b) shows the effect when the input pins are swapped ($B \leftrightarrow C$). Due to addition's commutative property, the output of both setups remains the same. However, the set of PMOS transistors experiencing NBTI stress is changed. As we can see in Figure 2(b), transistor T1 is now under stress since its gate input is '0', showing the effect of this simple chnage. An attacker can use this observation to specifically target a subset of transistors in full adders of a multiplier (and scale it up to MAC units of AI accelerator), causing accelerated aging to induce failure.

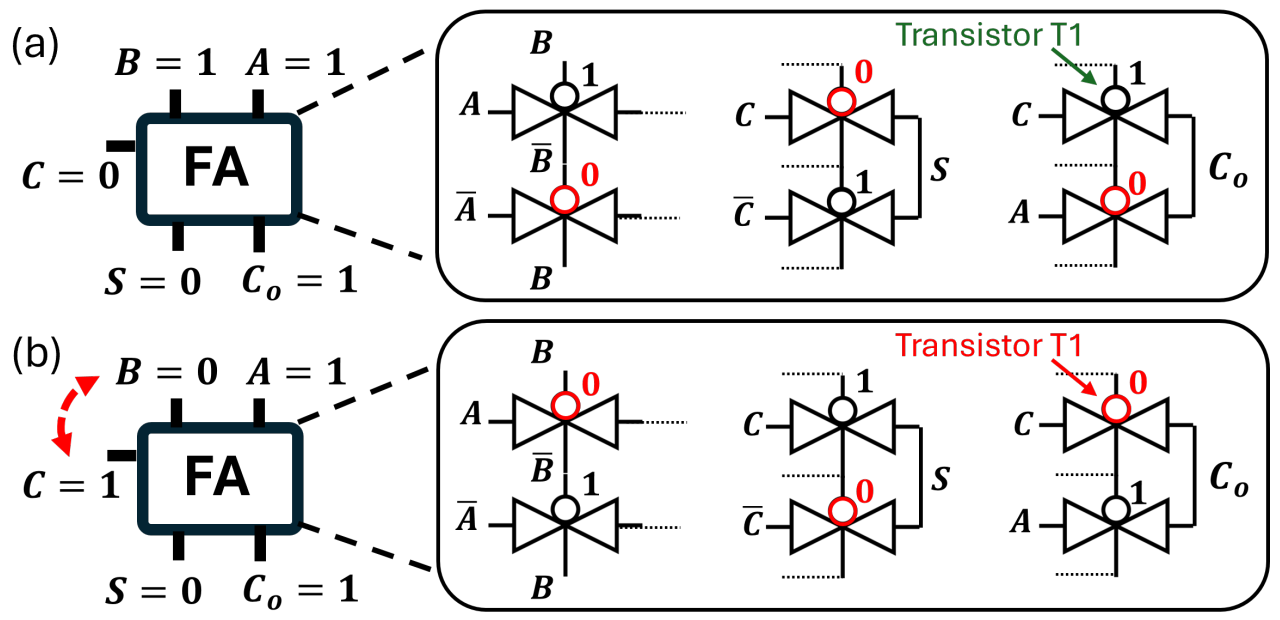

**Figure 2.** Effect of the attack on a full adder: (a) full adder with input '110', showing no stress on transistor $T1$; (b) full adder with swapped input pins ('101'), causing shifted stress on transistor $T1$.

### *B. Attacking a MAC unit*

Next, we show how the tampering can be scaled up and implemented in a MAC unit of an AI accelerator (like a systolic array) to induce errors. A typical MAC unit consists of a multiplier and adder along with supporting peripheral circuits as shown in Figure 3(a). For demonstration purposes, we consider an Array multiplier and a transmission gate based full and half adders noting that the attack is applicable for other multiplier/adder architectures as well. Since real world inputs are random, we represent the stress on transistors as the probability that its gate input is zero.

Implementing the proposed attack in a multiplier (or any complex circuit) that consists of multiple commutative blocks (full/half adders here), raises two questions (a) *which blocks (adders) to attack/tamper?* and (b) *how should the inputs of each block be rearranged?* An attacker may tamper all adders in a multiplier. However, this reduces the stealthiness of the attack as it increases the probability of getting spotted upon visual inspection. Moreover, the delay is determined by the critical path(s). Hence, tampering all adders may be an overkill.

Algorithm-1 presents a greedy depth-first graph traversal strategy to identify the weakest paths and adders that can be tampered for maximum possible damage at design-time. Algorithm-1 is agnostic to the choice of ML model and only requires the following inputs: the circuit to tamper (MAC unit of a systolic array here), probability that circuit's primary input bit is '0' and the number of adders that we want to target per path (i.e., attack budget) that is provided by attacker.

As shown in Algorithm-1, the attack starts by identifying top-k critical path(s) based on each path's delay. Here, we consider multiple paths for tampering to account for process variation and randomness in input, which can result in unpredictable $V_{th}$ change for transistors in different paths [38]. As a result, it is possible that some paths are more vulnerable to the attack than the critical path. Next, the goal is to identify which adders to target in chosen path for maximum effect.

**Algorithm 1** Rewiring the MAC units in AI accelerators

**Input:** Circuit $C$ (MAC unit here), $\alpha_{in}[.]$ probability of primary inputs being zero, $N_{target}[.]$ number of blocks (Full Adders here) that will be targeted per path $p$

```
1:  Crit[.] = get top-k critical paths //Cadence simulation
2:
3:  FOR path p ∈ Crit:
      Reset Circuit C
      FOR n = 0 to N_target[p]:
4:      FOR adder g IN path p:
5:        d[0] = SIMULATE(α_in, C)
6:          FOR w_i ∈ all input combinations to adder:
7:            d_w = SIMULATE(α_in, w_i)
8:            ind = argmax_w(d[w] − d[0])
            C = Apply(W_ind to g)
9:  𝔻_C[p] = SIMULATE(α_in, C)
10: Return top − n from Sort(𝔻_c[p])
```

For this purpose, Algorithm-1 starts a depth first graph traversal. Assuming that the circuit (multiplier here) is a graph, each path can be envisioned as branches of the graph, while the adders are the nodes. The algorithm starts by iterating through the candidate critical paths identified in the circuit. For the $p^{th}$ path it iterates for $N_{target}[p]$ number of times (Line 3-5), where $N_{target}$ is an attacker provided attack budget. For each adder in a path, we evaluate every possible input combination (Line 6-7). For a full adder, there are only $3! = 6$ different ways to apply the three inputs. Hence, we do an exhaustive evaluation simulating the aging behavior and corresponding delay for each input combination considering input probabilities via a combination of Cadence simulation and aging models (Equation (2)). For each adder, we note the max stress achievable and apply the corresponding tampering. The worst tampering identified in the current path is then applied following a greedy strategy. This process is repeated for $N_{target}$ number of adders per path to identify the most vulnerable adders. In each iteration, we identify a new adder and its corresponding tampering, following the greedy strategy.

The process is repeated for each path separately. Once all paths have been analyzed, we sort the corresponding delay in line 10, eventually picking the most sensitive path(s) with the most impactful tampering. This structured approach enables efficient, targeted attacks with minimal rewiring effort while maximizing long term circuit degradation.

Depending on the attacker's objective and stealth constraints, the value of $N_{target}$ can be varied. This control allows for flexibility in attack strength versus detectability. Rewiring fewer gates improves stealth, while a broader attack can inflict more severe aging effects.

### *C. Final Implementation Considering an AI Accelerator*

Finally, Figure 3 shows the overall implementation considering a systolic arrray-based AI accelerator. These accelerators are widely adoppted in machine learning inference and training (e.g. Google's TPU), and rely heavily on MAC operations as their computational backbone. By integrating our attack into their architecture, we demonstrate its practical impact on real world systems.

Systolic arrays are composed of a structured 2D grid of processing elements (PEs), where each PE is resposible for executing MAC operations. A MAC operation consists of multiplication followed by an accumulation operation as shown in Figure 3(a). The multiplier can be either in fixed point or floating point format. Typically $D \times D$ systolic array performs matrix multiplication by streaming one operand matrix $X$ row wise and the other opeand matrix $W$ column wise. Each PE receives a pair of inputs $(X_r, W_c)$, performs a multiplication, and accumulates the result with previous partial sums. The architecture benefits from localized data movement and reduces memory bottlenecks, achieving better throughput and energy.

For integer and fixed point multiplication, the multiplier architecture is relatively straightforward. To implement the attack, we identify and tamper chosen full adders in multipliers in side each MAC unit following Algorithm-1. However, floating-point multiplication differs slightly from integer or fixed-point multiplication. For example, in an IEEE 754 single-precision (float32) multiplier, a number consists of three components: a 1-bit sign ($S$), an 8-bit exponent ($E$), and a 23-bit mantissa ($M$). Figure 3(b) shows the multiplication of two float32 numbers $P \times Q$. As we can see in Figure 3(b), the multiplication involves XOR-ing the sign bits, adding the exponents and multiplying the mantissa, which is equivalent to integer multiplication. Hence, we can implement the attack in the mantissa multiplication for floating point multpliers; the idea can be adopted for other bit-precisions like 12-bit and 8-bit floating-point representations as well. Since floating point is the default choice for most AI models, we assume floating point multipliers in the systolic array for demonstration purpose.

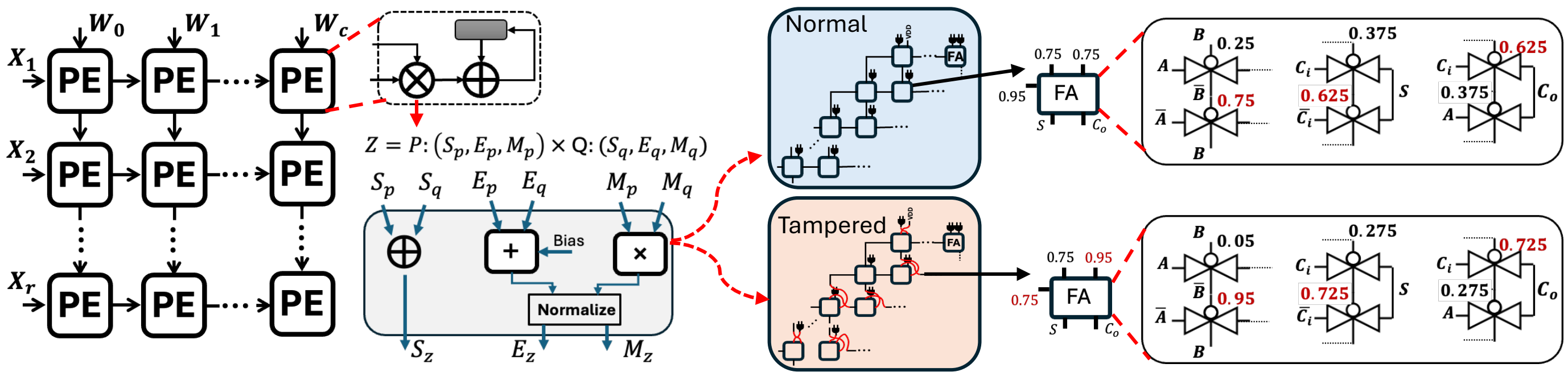

**Figure 3.** (a) Systolic array architecture with embedded MAC units. (b) steps of floating-point multiplication. (c) array multiplier inputs before and after the attack. (d) Prob. that input is '0' ($\alpha$) in full adder and (c) each transistor's $\alpha$ before and after applying the proposed attack.

Figure 3(c) illustrates the attacked architecture, where some FAs in the multiplier have been selectively rewired, highlighted in red. As we can see, the overall structure and functionality of the systolic array remains unchanged despite the attack. The only modification involves reordering the input connections within the adders of the multiplier. This targeted rewiring does not introduce any area overhead, as no new hardware is added; it only permutes exisiting inputs in commutative logic. Figure 3(d) shows the eventual effect of this tampering in a full adder with hypothetical probabilities that its input A,B,C are 0. Here, we use a probabilistic notation since real-world inputs can be random. Figure 3(e) also shows the corresponding probabilities that the input to PMOS is 0 (under stress); the probabilities can be obtained via Cadence simulation. As we can see in Figure 3(e), the tampering results in an increase in stress for some transistors, making them more likely to fail.

## V. Experimental Result

In this section, we begin by detailing the simulation setup, followed by evidence to support the proposed method.

### A. Experimental Setup

For evaluation of the various hardware e.g., multipliers, we used the electronic design automation (EDA) tool Cadence Spectre and OCEAN. The simulation was conducted using GlobalFoundries' 22nm (GF22nm) process technology. The supply voltage ($V_{dd}$) in GF22nm is $0.8V$, and the nominal threshold voltage of all PMOS at $t = 0$ is $V_{th,nominal} = 0.45V$ (without process variation). We use Cadence to determine the delay and error likelihood for various configurations. To evaluate the proposed approach, we conducted experiments on 6-bit, 8-bit, 10-bit and 16-bit multipliers. We also consider two types of multipliers, namely the Array and Wallace Tree Multiplier to thoroughly evaluate the proposed method across various bit widths and architecture.

For transistor degradation, we consider the NBTI process, noting that other aging mechanisms like HCI and TDDB can also be incorporated; we leave this exploration for future work. The change in $V_{th}(t)$ is modeled using Equation (2). To model process variation, we assume Gaussian distribution using Equation (3) with standard deviation of $\sigma_{V_{th}} = 0.02$. We perform Monte Carlo simulations with 50k iterations to thoroughly evaluate the proposed method under various process variation scenarios. In each iteration, we vary the initial threshold voltage $V_{th}(t = 0)$ of the PMOS transistor randomly following Equation (4). We also compare our attack with prior work, namely workload aware lifetime trojan [25] and process reliability based trojan [24].

Finally, we demonstrate the effect of the attack on error likelihood and accuracy of AI models. For this purpose, we use a mix of AI models like CNNs, GNN, and transformer, namely Darknet, Resnet18, VGG19, VGG11, DGCNN and BERT models for comprehensive evaluation [39], [40]. We also use various datasets including Cifar-10 for CNNs, NCI-1 for GNN and SST-2 for the BERT model where NCI-1 involves graph classification of chemical compounds and SST-2 is a benchmark for sentence level sentiment analysis.

### B. Effect of Proposed Attack on a Single Multiplier

First, we evaluate the impact of the proposed attack on the stress distribution in an Array multiplier. Figure 4 shows the percentage of PMOS transistors as a function of the stress they experience. As mentioned earlier, stress is quantified by the probability that gate input of a PMOS is '0' ($\alpha$). For this experiment, we consider two extremes: no tampering and full-tampering, where every full adder in the multiplier is tampered greedily following Algorithm-1. Next, we apply all $2^{2n}$ inputs where $n$ is the bit width of the multiplier and count how often each PMOS experiences negative bias i.e., $V_{gs} < 0$. The resulting distribution is shown in Figure 4. Before tampering, we see that most transistors exhibit a near equal likelihood that $V_{gs} < 0$. However, after the tampering the $\alpha$ values of transistors are pushed towards both extremes. Some PMOS now experience negative bias more frequently than before (high $\alpha$), creating bottlenecks. These PMOS degrade sooner than normal, resulting in incorrect computations.

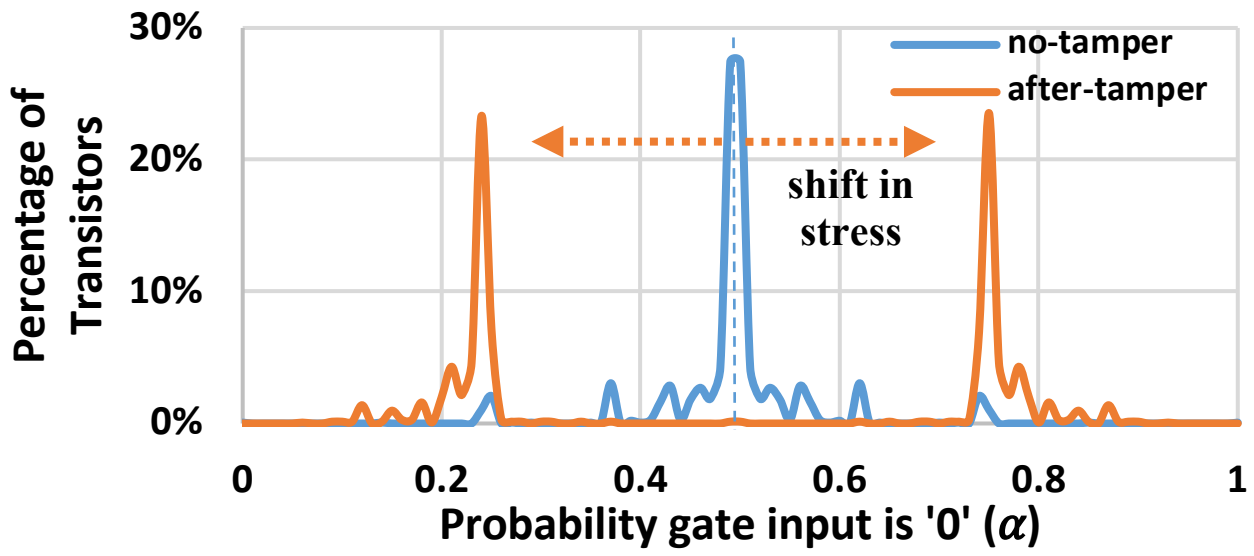

**Figure 4.** Stress distribution, measured as probability of gate input being '0' across all transistors in an array multiplier

Next, we show the impact of this distribution shift on the propagation delay of both 8-bit Array and Wallace Tree multiplier, assuming no process variation and equal probability of multiplier inputs being zero/one (i.e., $\alpha = 0.5$). For this experiment, we only tamper the most critical path (i.e., one path only); we extend the experiment with multiple paths and with process variation later. Figure 5 illustrates the accelerated degradation when different number of Full Adders along the critical path of both multipliers are tampered using Algorithm-1. We refer to the configurations as M-x-y, which indicates that x number of paths, and y% of adders in each path have been tampered. All delay values are normalized to the delay of the 8-bit array multiplier at $t = 0$. For comparison, we also show the untampered case M-0-0, which experiences natural aging.

As we can see in Figure 5(a), for 8-bit array multiplier, tampering all adders in the most critical path (M-1-100%) attack results in an additional 16% and 29% delay after one and four years respectively, compared to natural aging in M-0-0. Similarly, as shown in Figure 5(b), the M-1-100% attack in Wallace Tree multiplier results in an additional 29% and 48% increase in delay compared to the baseline (M-0-0).

To make the attack stealthier, attacker can also tamper only a subset of full adders within the critical path instead of modifying all of them, which we also consider here. For this purpose, we rank the commutative blocks on the critical path following Algorithm-1, and then select the most vulnerable ones for tampering. We evaluate three partial tampering levels: top-25% (M-1-25%), top-50% (M-1-50%), and top-75% (M-1-75%) of the most vulnerable full adders. As shown in Figure 5(a), these configurations in the 8-bit array multiplier increase the aging delay by 10%, 19%, and 25%, respectively, compared to the baseline (M-0-0) after four years of operation. For reference, tampering all critical-path adders (M-1-100%) produces a 29% increase. A similar trend is observed for the 8-bit Wallace Tree multiplier in Figure 5(b), where M-1-25%, M-1-50%, and M-1-75% tampering cause 19%, 39%, and 45% delay increases, respectively, while full tampering (M-1-100%) results in a 48% increase. The increase in delay is a result of the shift in stress distribution (as shown in Figure 4). Transistors experiencing very high stress degrade fast (i.e. $V_{th}$ increase over time), resulting in high delay.

Overall, Figure 5 shows that the simple rewiring following Algorithm-1 is an effective attack that can result in accelerated degradation in multipliers, resulting in timing violations and mispredictions during inferencing as we show later.

### *C. Impact of Multiplier Bit Width on Aging Attack*

To evaluate how proposed rewiring attack scales with circuit size, we apply the attack to 6-bit, 8-bit, 10-bit and 16-bit versions of both Array and Wallace Tree multipliers under similar conditions as Figure 5. For a fair comparison, we assume that the attacker only tampers 10% of the total number of adders in each multiplier, starting with the most critical path. For comparison, we evaluate the difference in delay between the tampered and baseline (M-0-0) delays, which is then normalized by the baseline delay at $t = 0$.

As we can see in Figure 6, higher bit width multipliers experience a significantly greater increase in aging induced delay due to the attack. Moreover, the relative delay between tampered and non-tampered multiplier delay grows with time and circuit size. For example, the 16-bit array multiplier exhibits a 11% increase in delay after one year compared to M-0-0 and a 20% increase after four years. However, the 6-bit multiplier has only 5% and 9% increase in delay in the same period. Similarly for 16-bit Wallace Tree multiplier we have a delay increase of 30% after one year compared to no-tamper, and 49% increase in four years, while the 6-bit multiplier has 16% and 25% increase respectively. These results show that the attack is more potent and scales with the bit width of the multiplier, even when the same fraction of adders in the multipliers are tampered.

### *D. Effect of Attack on Delay under Realistic Condition*

Thus far, we have considered no process variation or randomness in input. Process variation is modeled here as unequal threshold voltage ($V_{th}$) for each transistor. Similarly, randomness in inputs may cause unequal aging among the different paths in the multiplier. Hence, it is likely that other paths can also become a bottleneck due to randomness. To evaluate the impact of process variation and random input on the effectiveness of our proposed aging attack, we conduct a Monte Carlo simulation with 50k iterations using both 8-bit array and 8-bit Wallace Tree multipliers. In each iteration, we change the initial threshold voltage $V_{th}(t=0)$ of each transistor, assuming Gaussian distribution using Equation (4). We also assign a random value for $\alpha$ drawn uniformly from 0.1 to 0.9 capturing a broad range of signal behavior, from frequently high to mostly idle lines.

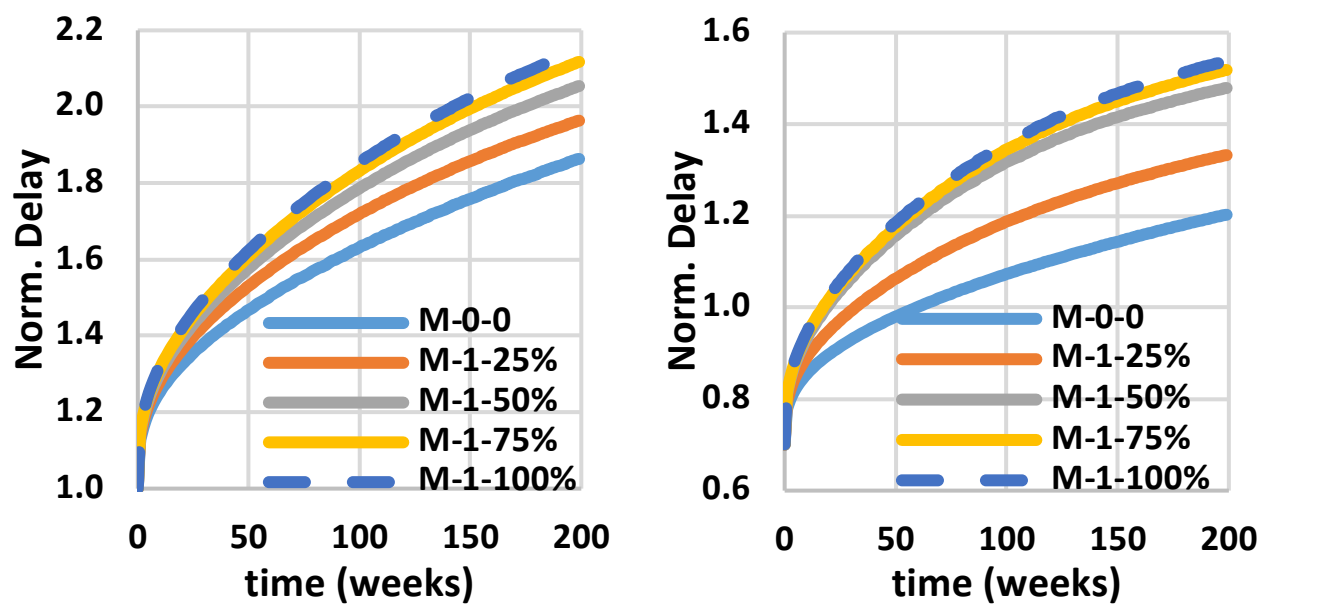

(a) (b)

**Figure 5.** Normalized delay over time for (a) 8-bit Array multiplier and (b) 8-bit Wallace Tree multiplier under different level of tampering assuming no randomness.

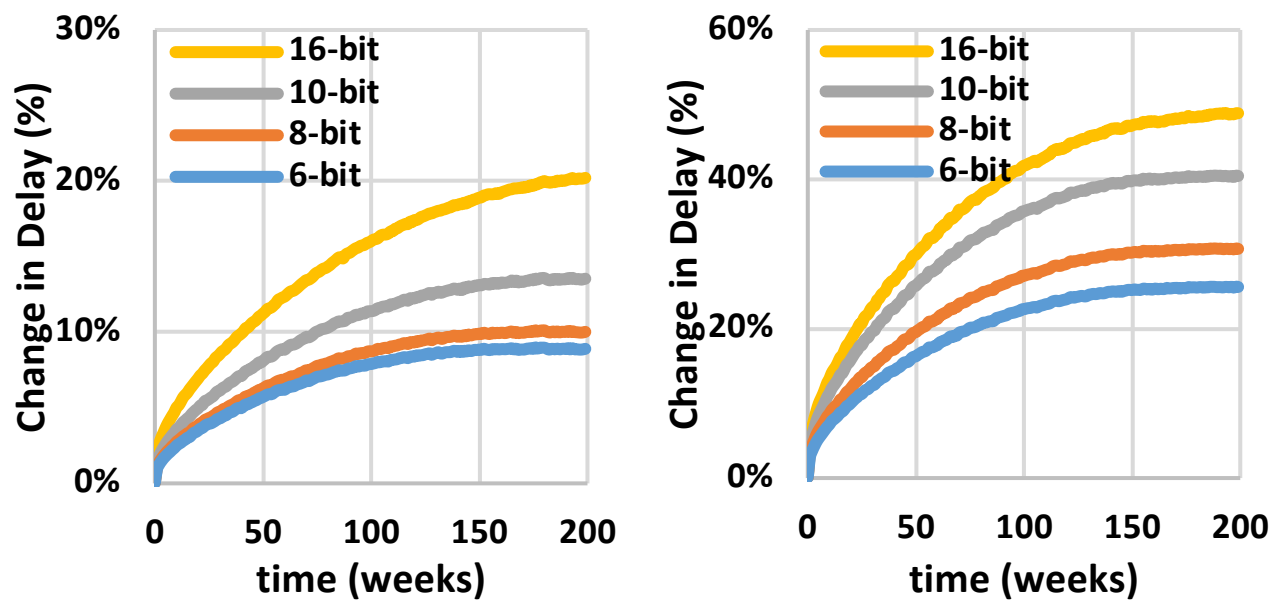

(a) (b)

**Figure 6.** Change in delay after tampering (M-1-100%) compared to normal aging for different bit widths in (a) Array multiplier and (b) Wallace Tree multipliers, over time assuming no randomness.

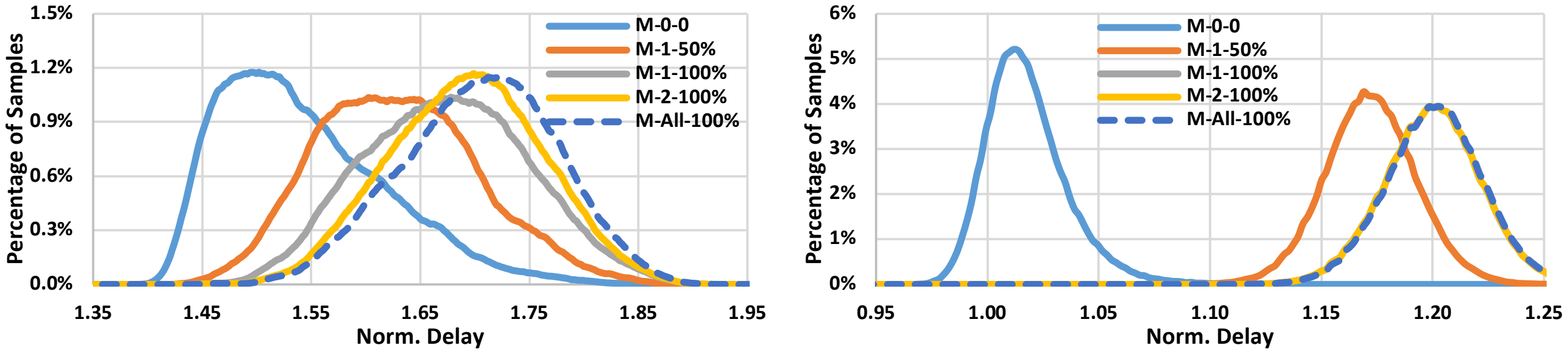

(a) (b)

**Figure 7.** Normalized *max* propagation delay based on Monte Carlo analysis after 1-year for (a) 8-bit Array multiplier and (b) 8-bit Wallace Tree multipliers under partial and full tampering, considering process and input variations.

Figure 7 shows the results across four tampering scenarios, (a) M-1-50%, (b) M-1-100%, (c) M-2-100%, and (d) M-All-100% after one year, compared against the baseline M-0-0 (natural aging). For the array multiplier, tampering increases aging by an additional 8%, 13%, 15%, and 16% for the four scenarios, respectively. Similarly, in the Wallace Tree multiplier, the four tampering cases result in 22%, 26%, 26%, and 26% additional degradation, respectively.

Figure 7 confirms that the proposed attack remains effective even under significant process variation and input variability, increasing the delay beyond the natural aging baseline. However, tampering multiple paths beyond the critical path (M-1-100%) produces only marginal delay increases. In other words, the maximum delay of the circuit quickly saturates after tampering the critical path, and targeting additional paths leads to little increase to max delay. This happens as the maximum delay is primarily dictated by the critical path. However, the trends are slightly different when we consider average delay of the multiplier considering different input combinations.

Figure 8 shows the average delay for both multipliers for various input combinations at different levels of tempering after one year. For the 8-bit Array multiplier, the average delay increased by approximately 3%, 5%, 8%, and 14% under the M-1-50%, M-1-100%, M-2-100%, and M-All-100%, respectively, compared to the baseline (M-0-0). Similarly, in the 8-bit Wallace Tree multiplier, the average delay increased by about 0.5%, 2%, 4%, and 9% for the same attack configurations.

These results show that additional tampering beyond the critical path contributes to a higher average propagation delay. Even though the max delay is relatively unchanged, the increase in average delay leads to some interesting effect on error likelihoods as we discuss next.

*E. Effect of Attack on Computation Accuracy of MAC Unit*

As the adders and multiplier degrades, propagation delay increases. This increases the likelihood of timing violations; Hardware manufacturers typically account for natural aging by including guard bands based on the worst-case aging of the circuit in its lifetime. However, these provisions are typically calibrated for standard aging and may fail due to the accelerated aging caused by the attack. The attack results in delays that exceed manufacturer's expectations, potentially resulting in timing violations, even after guard band. For instance, assuming a static guard band considering the worst case of natural aging (M-0-0) in Figure 8, we see that tampering multiple paths results in a larger number of timing violations even when the max delay is unchanged as we see in Figure 7.

Figure 9 shows the average error likelihood for the tampered multiplier in different bit widths. For this experiment, we assume that the manufacturer considered a 4-year worst case natural aging as baseline for guard band. To calculate error likelihood, we apply all $2^{2n}$ input combinations in a multiplier, where $n$ is the bit-width, and note the corresponding delay. Let worst-case propagation delay of multiplier with no tampering (M-0-0) after 4-years be $P_d$ (which we set as guard band limit) and let propagation delay for the tampered multiplier at time $t$ for input $(A, B)$ be represented as $P_d'(t)$. If for any input combination $(A, B)$, the delay $P_d'(t) > P_d$, it is likely to cause timing violations, leading to incorrect results (i.e. an error). We count the number of such occurrences via Cadence simulations.

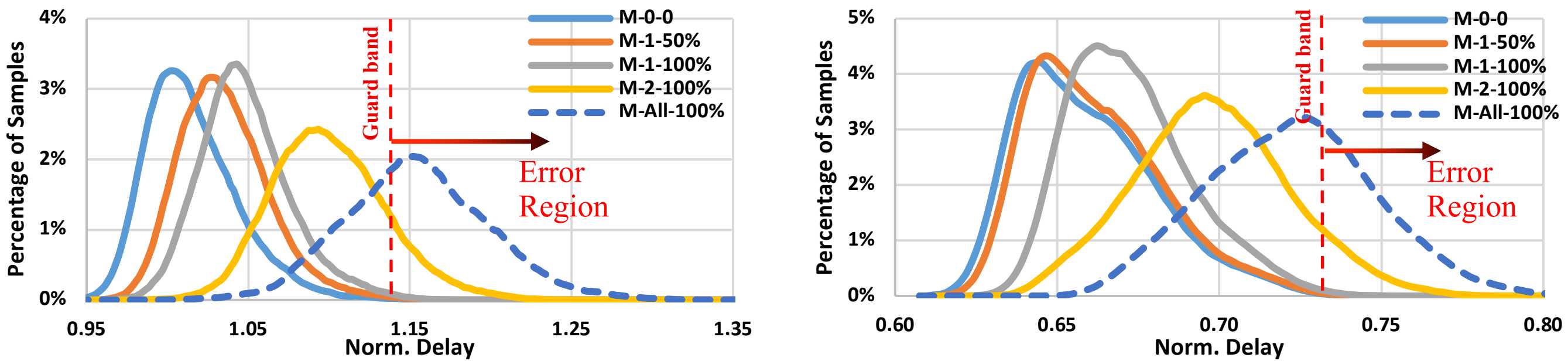

(a) (b)

**Figure 8.** Normalized *average* propagation delay considering Monte Carlo analysis after 1-year for (a) 8-bit Array multiplier and (b) 8-bit Wallace Tree multipliers under partial and full tampering, considering process and input variations.

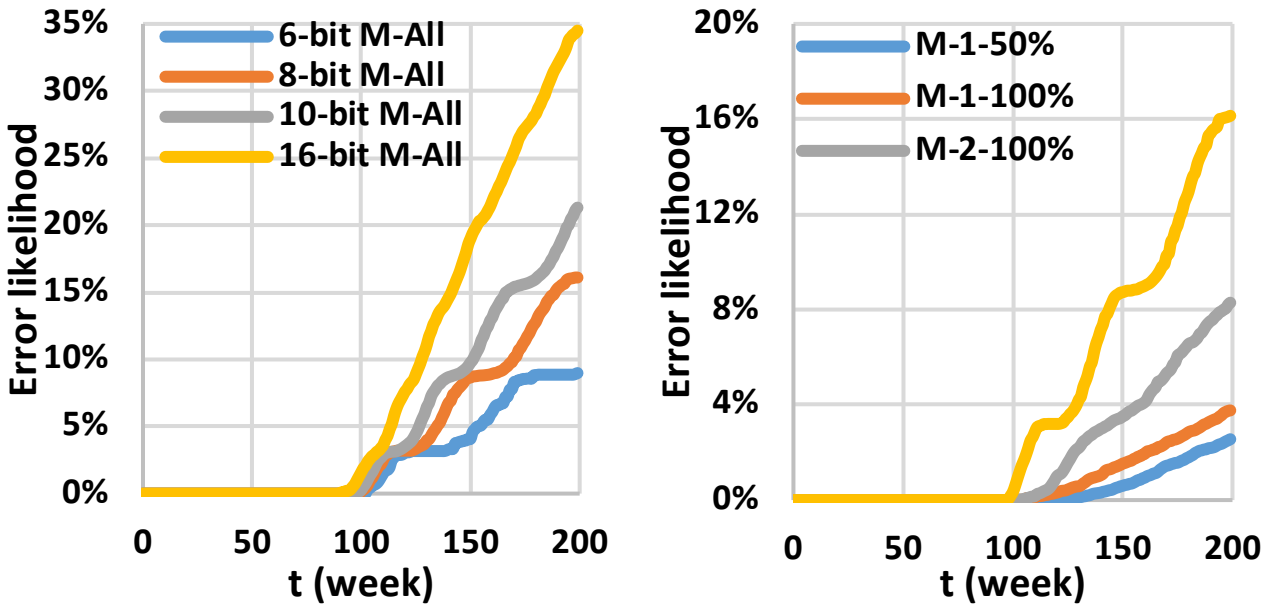

(a) (b)

**Figure 9.** Error likelihood over time for (a) array multiplier with different bit widths under M-All-100% tampering and (b) 8-bit array multiplier with different level of tampering.

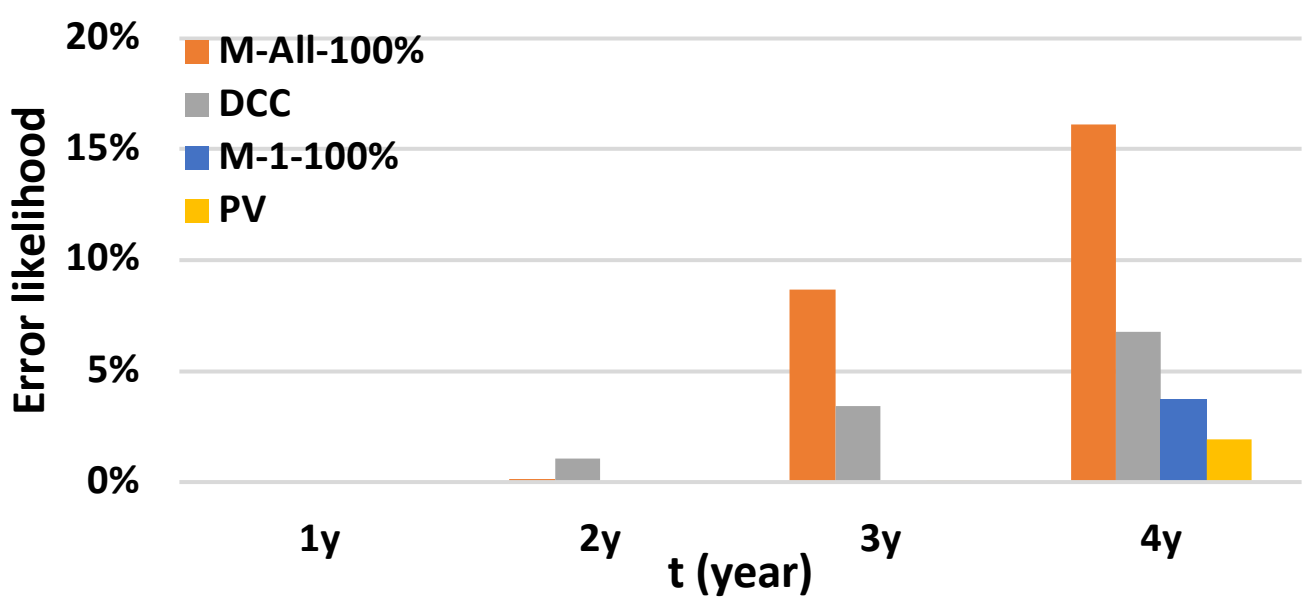

**Figure 10.** Comparison of error likelihoods during multiplication for workload aware lifetime trojan attack (DCC) and process reliability based trojan attack (PV), and two variants of the proposed attack (M-1-100% and M-All-100%) over a 4-years period.

Figure 9 shows the percentage of inputs that results in an error considering all the possible inputs; we refer this as 'error likelihood' here. For this experiment, we evaluate the M-All-100% for Array multiplier with different bit widths in Figure 9(a) and different levels of tampering for 8-bit Array multiplier in Figure 9(b). As illustrated in Figure 9(a), larger multiplier exhibits higher error likelihoods over time. For example, after four years, the 16-bit array multiplier reaches an error likelihood of 35%, compared to 16% for the 8-bit version in M-All-100% tampering.

The results in Figure 9(b) show that tampering more paths leads to more errors with 16% error likelihood in M-All-100% compared to 3% error likelihood in M-1-100%. This shows that while selectively tampering the critical path is effective in increasing the propagation delay, its impact on error likelihood is limited. Since inputs traverse multiple paths in a multiplier, tampering more paths ensures a higher chance of timing mismatches and violations, making this approach more effective in inducing errors. Moreover, randomness from process variation and input patterns can also result in other paths contributing to timing violations as we have shown in Figure 8. Wallace Tree architecture shows the same trend as array multiplier. We examine how this methodology translates to AI inference accuracy, and how its effectiveness differs in real world application scenarios in a later section.

### *F. Comparison with Other Attacks*

To assess the effectiveness of the proposed attack, we compare the proposed method's error likelihood over time with two closely related prior works: (a) Workload-Aware Lifetime Trojan [25], and (b) Process Reliability-Based Trojan [24]. The workload-aware trojan approach utilizes duty cycle correctors (DCCs) before and after the critical path to selectively accelerate aging in one part of the circuit while decelerating it in another, creating timing mismatches as different regions degrade at different rates. The process reliability based trojan introduces malicious process variations during fabrication, resulting in long term reliability degradation, in our case, we simulate this behavior by introducing 10% additional $V_{th}$ at $t = 0$ for transistors. Here, we choose these two baselines as they are most closely related to our approach. Other aging attacks like MAGIC [5] rely on input tampering, which does not follow our threat model of circuit tampering. Moreover, tampering inputs directly in an AI accelerator can result in immediate accuracy loss during inferencing, which is not stealthy.

Figure 10 shows the results of the various attacks. As we can see in Figure 10, tampering just the critical path (M-1-100%) outperforms process reliability-based Trojan method, but it causes fewer errors than workload-aware lifetime Trojan. However, tampering all paths (M-All-100%) outperforms both the workload-aware lifetime Trojan (referred as DCC in graph) and process reliability-based Trojan (PV) in terms of errors, demonstrating the attack efficacy. After four years of operation, the error likelihoods for the workload-aware lifetime Trojan, and process reliability-based Trojan are 7%, and 2%, respectively, while the proposed method achieves up to 16%.

In terms of implementation overhead, the proposed method introduces virtually no overhead since it merely rewires the input connections. The workload-aware lifetime trojan requires the insertion of duty cycle correctors, leading to area, and power overhead. The process reliability-based Trojan also does not incur implementation overheads as it relies on introducing high variability in transistor $V_{th}$. However, this is achieved via selective process-level modifications such as dopant engineering or gate-oxide tuning, which is more complex than simple reqiring as in proposed attack. Overall, the proposed attack introduces virtually no overhead, while causing comparable (or more) errors than prior work.

### *G. Effect of Attack on AI model Accuracy*

Next, we evaluate the impact of the errors (due to attack) on AI model inference. Since multipliers are core components of MAC units, any induced error within them directly affects the numerical computations in machine learning models. Specifically, timing violations within the multiplier circuits can cause incorrect partial products, which in turn propagate as noise through subsequent network layers.

As described before, we implement the attack at the floating-point mantissa multiplication stage; the sign and exponent fields remain untouched, preserving the overall range and direction of the numbers but introducing errors in mantissa. To assess the impact on end-to-end inference, we run experiments on a mix of neural network architectures including CNNs: Darknet, SqueezeNet, ResNet, VGG11, and VGG19 on the CIFAR-10 and Cifar-100 dataset, GNN on the NCI-1 dataset, and BERT Transformer on the SST-2 dataset. NCI-1 involves graph classification of chemical compounds and SST-2 is a benchmark for sentence level sentiment analysis.

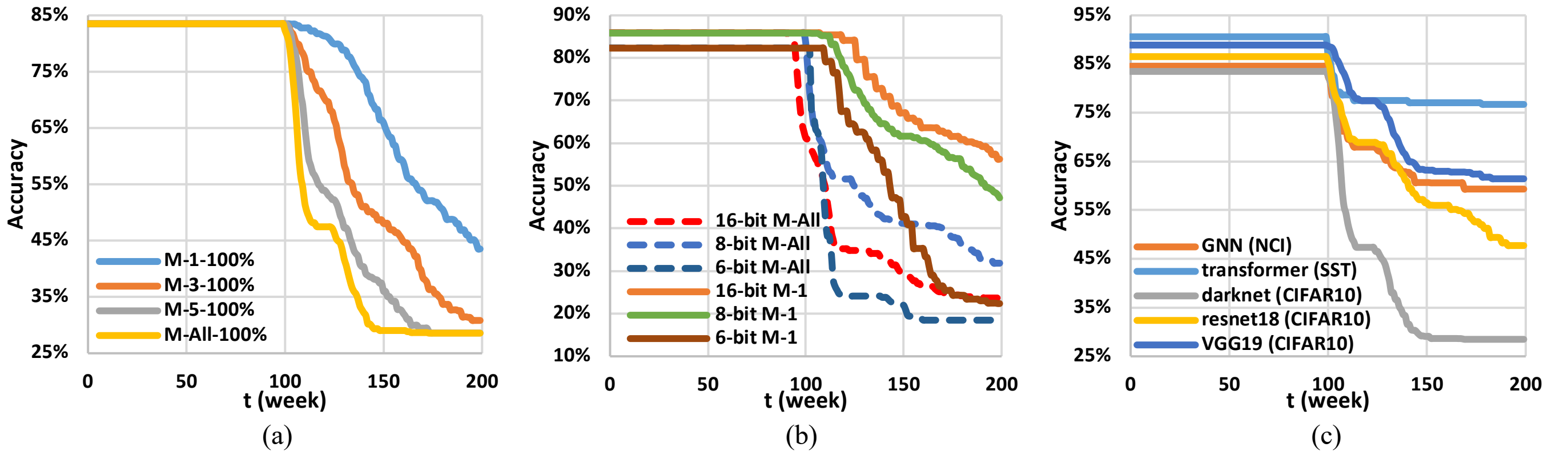

**Figure 11.** (a) Accuracy of Darknet with CIFAR-10 with 8-bit multipliers under varying tampering levels, (a) Accuracy of VGG11 with CIFAR-10 considering 6-bit, 8-bit, and 16-bit multipliers under M-All-100% and M-1-100% tampering, (b) Accuracy comparison of different models and datasets (CNNs, GNNs, transformers) with 8-bit multipliers under M-All-100% tampering.

As illustrated in Figure 11, model accuracy remains relatively constant in the initial years. This is due to the use of a timing guard band. Even though the attack causes degradation, the effect is not evident due to the guard band. Once the attack degrades the multiplier such that its latency exceeds the guard band, there is consistent degradation of accuracy over time regardless of the architecture or multiplier bit width.

In Figure 11(a), we present accuracy results for different tampering levels, assuming Darknet CNN with CIFAR-10 dataset and 8-bit Array multiplier as an example. As we can see in Figure 11(a), tampering more adders in the multiplier generally correlates with larger accuracy drops. Tampering only the critical path results in a 40% accuracy loss compared to $t = 0$, while M-3-100%, M-5-100%, and M-All-100% cause 53%, 55%, and 55% drop in accuracy, respectively. This happens as tampering more adders increases the chances of timing violations across multiple paths, resulting in higher number of errors (Figure 8).

Interestingly, the impact on accuracy diminishes as more blocks are tampered; for example, the difference between error likelihood under M-1-100% and M-3-100% tampering is 13%, whereas the gap between error likelihood under M-3-100% and M-All-100% tampering is just 2%, suggesting diminishing returns beyond a point. Moreover, while the degradation increases as more adders in the multiplier are tampered, tampering too many adders also increases the risk of detection during post-manufacturing tests. Hence, an attacker can choose the appropriate number of paths and adders to tamper (attack budget) based on desired impact and stealth requirements.

Next, in Figure 11(b), we present the accuracy of VGG11 as an example over time using multipliers of varying bit widths. We also consider two different tampering levels: only critical path (M-1-100%) and the entire multiplier (M-All-100%). From Figure 11(b), it is evident that smaller bit width multipliers not only exhibit lower initial accuracy but also experience a more pronounced decline. For instance, with M-1-100% tampering, a 6-bit multiplier's accuracy decreases from 82% to 22% over a four-year period, while a 16-bit multiplier's accuracy decreases from 86% to 56%. This can be attributed to the lower representation power in smaller bit widths, which results in lower accuracy to begin with. Moreover, the lower precision makes it more vulnerable to relatively small errors. Furthermore, as we can see in Figure 11(b) higher tampering levels cause more degradation always. In the case of the 16-bit multiplier, M-1-100% tampering results in a 30% drop in accuracy, whereas M-All-100% tampering leads to a more substantial reduction of 62%. This can be attributed to the higher error likelihood as we can see in Figure 9.

Finally, Figure 11(c) compares the accuracy degradation of different models and datasets using an 8-bit multiplier as an example with M-All-100% tampering. As we can see in Figure 11(c) there is accuracy loss across all models and datasets. For example, the transformer model exhibits a relatively modest 15% drop in accuracy, whereas Darknet experiences a substantial 55% loss. The lower accuracy drop in Transformer may be attributed to the use of a relatively simple SST-2 dataset (2 classes only). For more complex language datasets, the accuracy drop is expected to be higher. However, we see 38% accuracy drop on average for all models and datasets after 4-year, which is unacceptable for any real-world application.

## VI. Conclusion

In this work, we have presented a new hardware aging attack that compromises AI accelerators by tampering the adders within a multiplier. The attack shows that wiring choices in commutative blocks like adders can substantially impact the degradation level of circuits even when they are functionally equivalent. The proposed attack is evaluated on different multiplier architectures with varying bit widths under process variation and random input activity. Through Monte Carlo simulations, we have demonstrated that the attack not only increases the delay, but also the error likelihood of the multiplier. Since multiplication forms the backbone of AI applications, this attack can make AI accelerators dysfunctional within a short time despite including a worst-case timing guard band. Our experiments with systolic arrays show up to 64% accuracy loss after four years of operation.